\documentclass[amsmath,twocolumn,aps,prl,superscriptaddress]{revtex4-2}
\usepackage{graphicx}
\usepackage{dcolumn}
\usepackage{bm}
\usepackage{amsmath,amssymb}
\usepackage{color}
\usepackage{amsfonts}
\usepackage{bigints}
\usepackage{booktabs}
\usepackage{multirow}
\usepackage{float}
\usepackage{environ}
\NewEnviron{ignore}{}{}

\begin{document}

\title{Phonon hydrodynamic regimes in sapphire}

\author{Takuya~Kawabata}
\affiliation{Department of Physics, Gakushuin University, Tokyo 171-8588, Japan}

\author{Kosuke~Shimura}
\affiliation{Department of Physics, Gakushuin University, Tokyo 171-8588, Japan}

\author{Yuto~Ishii}
\affiliation{Department of Physics, Gakushuin University, Tokyo 171-8588, Japan}

\author{Minatsu~Koike}
\affiliation{Department of Physics, Gakushuin University, Tokyo 171-8588, Japan}

\author{Kentaro~Yoshida}
\affiliation{Department of Physics, Gakushuin University, Tokyo 171-8588, Japan}

\author{Shu~Yonehara}
\affiliation{Department of Physics, Gakushuin University, Tokyo 171-8588, Japan}

\author{Kohei~Yokoi}
\affiliation{Department of Physics, Gakushuin University, Tokyo 171-8588, Japan}

\author{Alaska~Subedi}
\affiliation{CPHT, CNRS, \'{E}cole polytechnique, Institut Polytechnique de Paris, 91120 Palaiseau, France}

\author{Kamran~Behnia}
\email{kamran.behnia@espci.fr}
\affiliation{Laboratoire de Physique et d'\'{E}tude des Mat\'{e}riaux (ESPCI—CNRS—Sorbonne Universit\'{e}), \\PSL Research University, Paris, 75005, France}

\author{Yo~Machida}
\email{yo.machida@gakushuin.ac.jp}
\affiliation{Department of Physics, Gakushuin University, Tokyo 171-8588, Japan}

\date{\today}

\begin{abstract}
When an ideal insulator is cooled, four regimes of thermal conductivity are expected to emerge one after another. Two of these, the Ziman and the Poiseuille, are hydrodynamic regimes in which collision among phonons are mostly Normal. It has been difficult to observe them, save for a few insulators with high levels of isotopic and chemical purity. Our thermal transport measurements, covering four decades of temperatures between 0.1 K and 900 K, reveal that sapphire displays all four regimes, despite its isotopic impurity. In the Ziman regime, the thermal conductivity exponentially increases attaining an amplitude as large as 35,000 W/Km. We show that the peak thermal conductivity of ultra-pure, simple insulators, including diamond, silicon and solid helium, is set by a universal scaling depending on isotropic purity. The thermal conductivity of sapphire is an order of magnitude higher than what is expected by this scaling. We argue that this may be caused by the proximity of optical and acoustic phonon modes, as a consequence of the large number of atoms in the primitive cell. 
\end{abstract}

\maketitle

In non-magnetic insulators, heat is exclusively transported by phonons. Thanks to their larger group velocity, acoustic modes usually dominate. In 1966, Guyer and Krumhansl \cite{Guyer1966} identified  four thermal transport regimes. Let us recall them one by one.

\begin{figure}[H]
\begin{center}
\centering
\includegraphics[width=8cm]{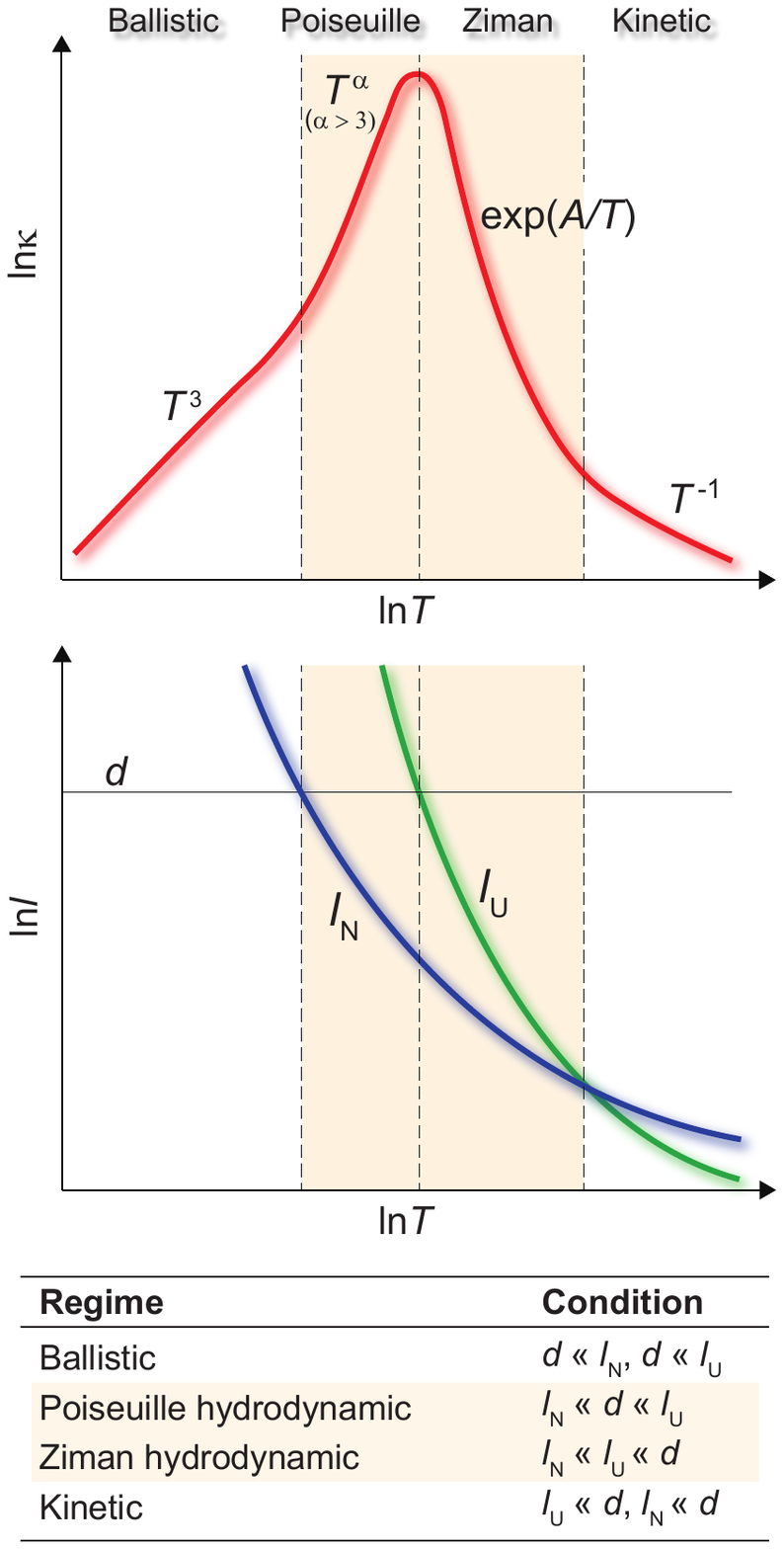}
\vspace*{-0.5 cm} 
\caption{Four regimes in temperature dependence of thermal conductivity. $l_{\rm U}$ and $l_{\rm N}$ represent the mean-free-path of Umklapp and Normal scattering, respectively. Hierarchy of $l_{\rm U}$, $l_{\rm N}$, and the sample dimension $d$ determines the four regimes. Dominant Normal collisions open the two hydrodynamic regimes in between the ballistic and kinetic regimes.}
\label{four}
\end{center}
\end{figure}

At temperatures exceeding the Debye temperature, the phonon wave-vector $k$ is comparable to the width of the Brillouin zone $G$. 
In this regime, called kinetic, scattering between two phonons can give rise to a third phonon with a wave-vector larger than $G$. Such an ``Umklapp'' event~\cite{Peierls} leads to loss of $\hbar G$ momentum and a resistance to the flow of heat. The average number of phonons increases linearly with $T$ and thermal conductivity $\kappa$ decreases linearly with warming $\propto T^{-1}$ (Fig.~\ref{four}(a)). The scattering rate in this limit is bounded by $\tau_P^{-1}=\frac{\hbar}{k_BT}$ \cite{Behniapl,Mousatov}. As the sample is cooled down, the wave-vector of phonons is reduced, following $k\simeq k_{\rm B}T/\hbar v_{\rm s}$. As a consequence, Umklapp processes rarefy, and instead momentum-conserving (Normal) collisions become dominant  (Fig.~\ref{four}(b)).  This leads to hydrodynamics, and the two regimes known as Ziman and Poiseuille. An  exponential increase of $\kappa$ in the Ziman hydrodynamic regime is driven by the decay of the phonon population capable of giving rise to Umklapp events. In the Poiseuille hydrodynamic regime, Normal collisions dominate and  Umklapp collisions become even less frequent than boundary collisions, this boosts heat conduction and leads to a $\kappa$ varying faster than $T^3$. Cooling further, phonons travel across the sample without meeting other phonons. Their wavelength becomes so long that the crystal wall becomes the exclusive scattering center. In this  ballistic (or Casimir~\cite{CASIMIR1938495}) regime, the thermal conductivity follows a $T^3$ temperature dependence. 
\begin{figure*}[t]
\begin{center}
\centering
\includegraphics[width=19cm]{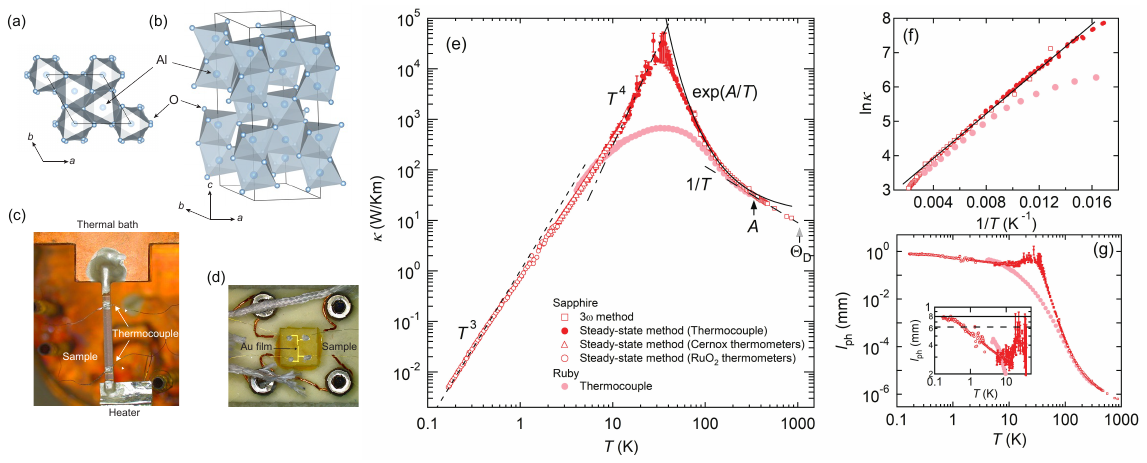}
\vspace*{-20cm} 
\caption{(a) Top view and (b) side view of crystal structure of sapphire. A thermal conductivity measurement setup for (c) steady-state method and (d) 3$\omega$ method. Thermal conductivity $\kappa$ of sapphire and ruby as a function of temperature. For sapphire, $\kappa$ displays an exponential increase ($\kappa\propto\exp(A/T)$ with $A=$ 337 K and forms a prominent peak around 40 K, below which $\kappa$ falls faster than $T^3$ as a signature of phonon Poiseuille flow. Temperature variation of $\kappa$ changes from $1/T$ to the exponential across $T=A$. An Arrhenius plot shown in the panel (f) clearly demonstrate the presence and absence of exponential behavior in sapphire and ruby, respectively. (g) A pronounce Poiseuille peak is detected in temperature dependence of phonon mean-free-path, which increases aiming to attain the crystal dimension at low temperature as shown in the inset of panel (g).}
\label{kappa}
\end{center}
\end{figure*}
These four regimes~\cite{Guyer1966,beck1974,Cepellotti2015} provide a road-map in the quest for detecting signatures of phonon hydrodynamics in the two regimes where Normal collision between phonons dominate: i.e. the Ziman and Poiseuille regimes~\cite{beck1974,Cepellotti2015,machida2024}.

The two hydrodynamic regimes are rarely observed in real insulators~\cite{Cahillcsi}, presumably because they are severely hindered by the presence of defects, and chemical or isotopic impurities. Recent studies, however, have found that exceptional purity is not the necessary condition for detecting a phonon Poiseuille flow. It has been observed in strontium titanate, a quantum paraelectric~\cite{Martelli}, in layered solids such as graphite and black phosphorus~\cite{MachidaBP,MachidaC}, and in semi-metallic antimony~\cite{Jaoui2022}

As for the Ziman heat transport regime, previous reports of its observation have been restricted to a few ultra-pure insulators composed of single-isotope elements, such as solid helium~\cite{Lawson,Thomlinson1972}, CsI~\cite{Cahillcsi}, and NaF~\cite{Jackson}. Surprisingly, the thermal conductivity of naturally available sapphire (Al$_2$O$_3$) displays the Ziman regime. While aluminum is pure ($^{27}$Al: 100~\%), this is not the case of oxygen ($^{16}$O: 99.76~\%, $^{17}$O: 0.04~\%, $^{18}$O: 0.21~\%). Nevertheless, in 1951, Berman~\cite{Berman1951} reported an exponential temperature dependence based on a limited number of data points (see Fig.~\ref{comp}). The issue has not been reexamined since then.

In this paper, we present a study of heat transport in sapphire, documenting its putative Ziman regime and Poiseuille regime. Prior to our study,  solid helium has been the only case in which both regimes have been observed. We find that naturally occurring sapphire stands out among insulators by displaying both hydrodynamic regimes and a record peak thermal conductivity at their boundary. The observed conductance exceeds what is expected from a universal plot  of dimensionless thermal conductivity as a function of isotopic impurities obeyed by other insulators. We argue that the large number of atoms per primitive cell and/or the proximity of the optical and acoustic phonon branches are the key ingredients for this robustness against isotopic disorder.
\begin{figure*}[t]
\begin{center}
\centering
\includegraphics[width=18cm]{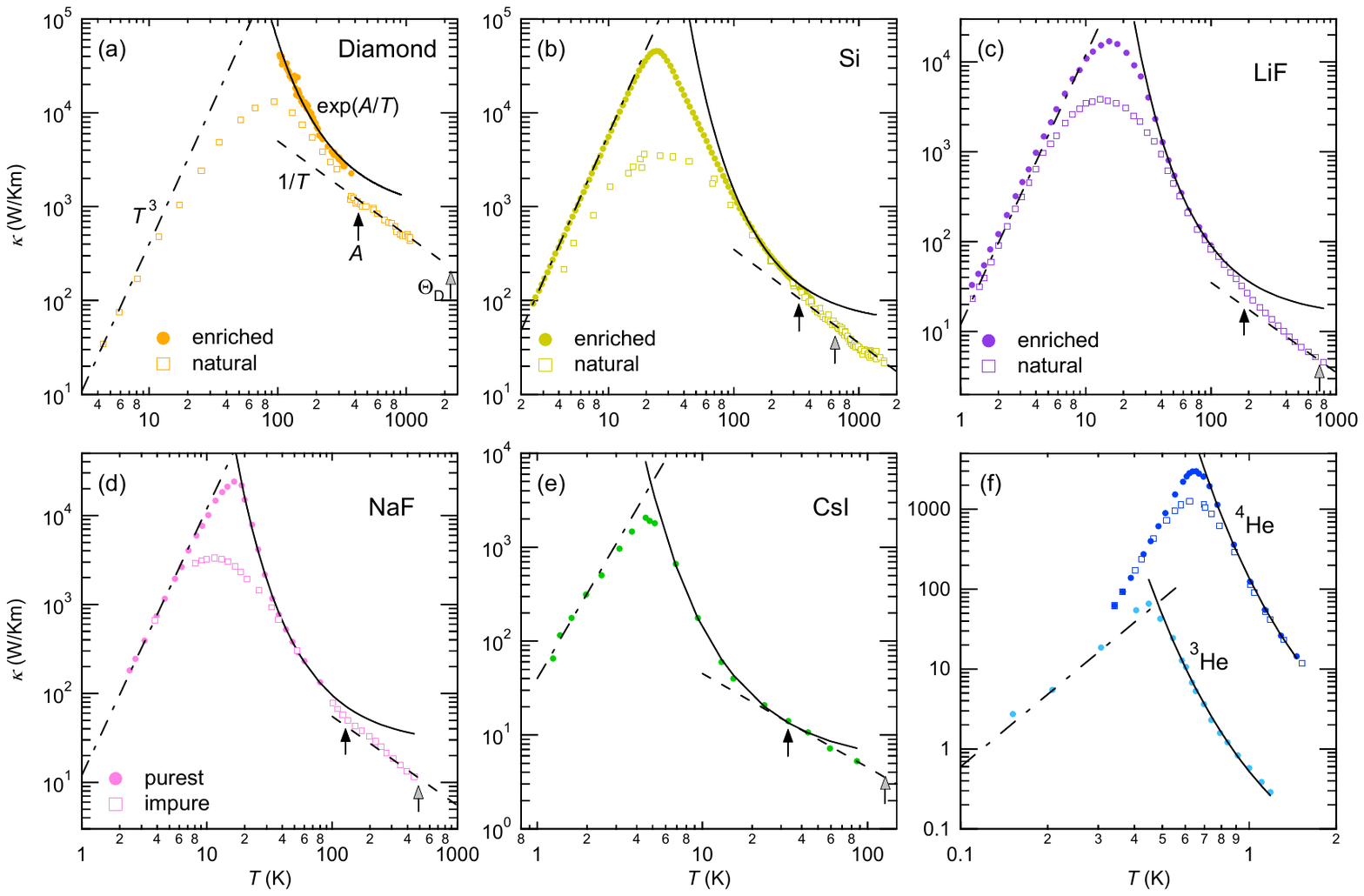}
\vspace*{-1cm} 
\caption{Temperature dependence of thermal conductivity of (a) diamond~\cite{BermanC,Olson,Wei}, (b) Si~\cite{Glassbrenner,Inyushkin}, (c) LiF~\cite{Slacknaf,Thacher}, (d) NaF~\cite{Slacknaf,Jackson}, (e) CsI~\cite{Cahillcsi}, and (f) $^4$He~\cite{Lawson} and $^3$He~\cite{Thomlinson1972}. The dashed-dotted, solid, and dashed lines represent the $T^3$, exponential, and $1/T$ dependence, respectively. The exponential behavior is fragile against the random isotopic mixture in diamond, Si, and LiF,  and impurities in NaF as shown in the panels (a)-(d). $A$ and Debye temperature $\Theta_{\rm D}$ of each material are denoted by closed and open arrows, respectively.}
\label{kappaval}
\end{center}
\end{figure*}

Sapphire has a layered trigonal structure with the space group $R\bar{3}c$. Each layer consists of edge-shared AlO$_6$ octahedra arranged in a honeycomb pattern (Fig.~\ref{kappa}(a)), and the layers are stacked such that each octahedron shares edges with the layers above and below it (Fig.~\ref{kappa}(b)).

Let us quantify the isotopic purity of sapphire. Isotopic disorder can be quantified by a dimensionless parameter~\cite{Inyushkin}
\begin{equation}
g=\sum_{i}f_i(\Delta M_i/M)^2.
\end{equation}
Here, $f_i$ is the concentration of the $i$-th isotope, whose mass $M_i$ differs from the average mass $M=\sum_{i}f_iM_i$ by $\Delta M_i=M_i-M$. In naturally abundant sapphire, that is in absence of any isotopic purification, $g\sim34\times 10^{-6}$. This is to be compared with that of the natural abundant diamond (74$\times 10^{-6}$) and silicon (200$\times 10^{-6}$). Thanks to the maturation of crystal growth technique, highly pure industrial sapphire single crystals with minimal contamination and defects are nowadays available.

The sapphire samples used in this study were purchased from Kyocera corporation. In order to acquire the thermal conductivity data in a wide temperature range, we employed the 3$\omega$ method and the standard steady-state method. The 3$\omega$ measurements were performed in the temperature range between 80~K and 900~K by utilizing the N$_2$ cryostat and the electrical furnace. The measurements, using a one-heater-two-thermometers steady-state method, were performed in the temperature range between 0.1~K and 300~K in a commercial measurement system (Quantum Design PPMS) and in a home-made dilution refrigerator.

Figure~\ref{kappa}(e) shows temperature dependence of thermal conductivity $\kappa$ of sapphire. As seen in the figure, there is a satisfactory match between the data obtained by the different methods in the different temperature ranges. The four conductivity regimes are all visible.  In the highest temperature regime, with decreasing temperature, $\kappa$ grows following  $1/T$. Upon further cooling, the temperature variation evolves into the exponential behavior $\kappa\propto\exp(A/T)$, signaling an entry to the Ziman regime. The result agrees with what was found by Berman~\cite{Berman1951} and with the first principle calculations~\cite{subedi21} (see Fig.~\ref{kyoob}(b)). The exponential increase persists down to 40 K, reaching a magnitude as large as 35,000 W/Km. Such a magnitude has only be attained in isotopically pure diamond~\cite{Wei} and Si~\cite{Inyushkin}  (see Figs.~\ref{kappaval}(a) and (b)). Below the peak, $\kappa$ falls rapidly following $\propto T^4$. Such a faster than $T^3$ decrease of $\kappa$ is a hallmark of the phonon Poiseuille flow~\cite{Mezhov,Thomlinson1969,Kopylov,Zholonko,Martelli,MachidaBP,MachidaC}. Further cooling the sample below 1 K eventually leads to the ballistic regime and a $T^3$ behavior.
\begin{figure*}[t]
\begin{center}
\centering
\includegraphics[width=17cm]{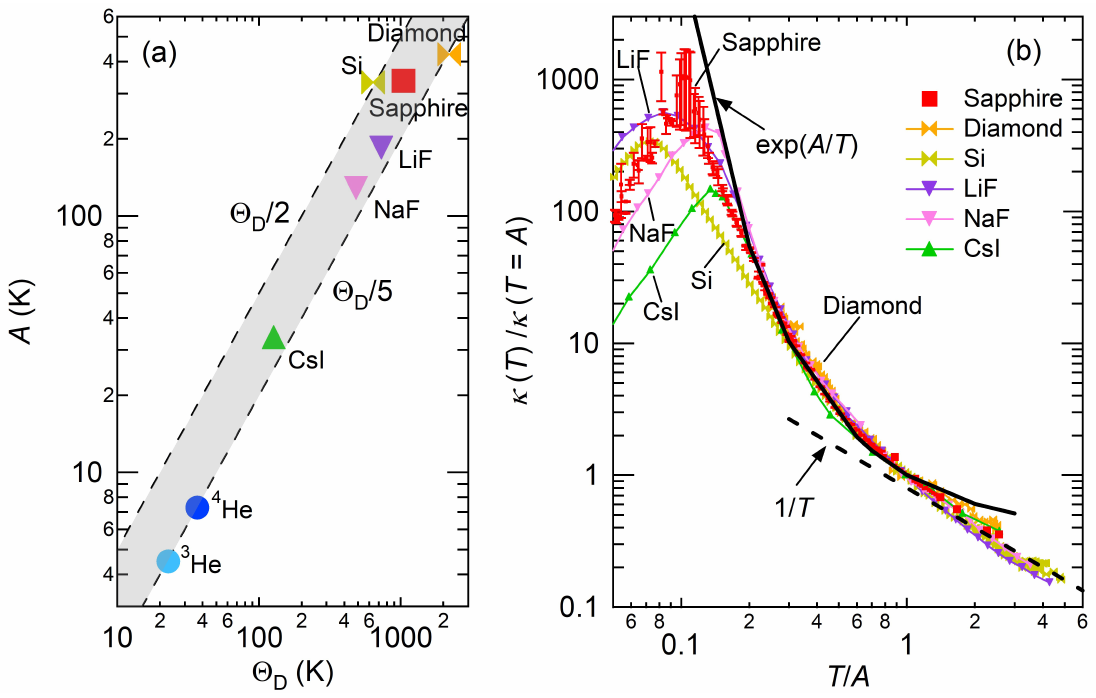}
\vspace*{-14cm} 
\caption{(a) Comparison of $A$ and $\Theta_{\rm D}$ in insulators showing the exponential behavior in thermal conductivity. Their ratio $A/\Theta_{\rm D}$ remains between 1/5 and 1/2. (b) Normalized $\kappa$ by those values at $T=A$ as a function of $T/A$.}
\label{A}
\end{center}
\end{figure*}

\begin{figure}[t]
\begin{center}
\centering
\includegraphics[width=12cm]{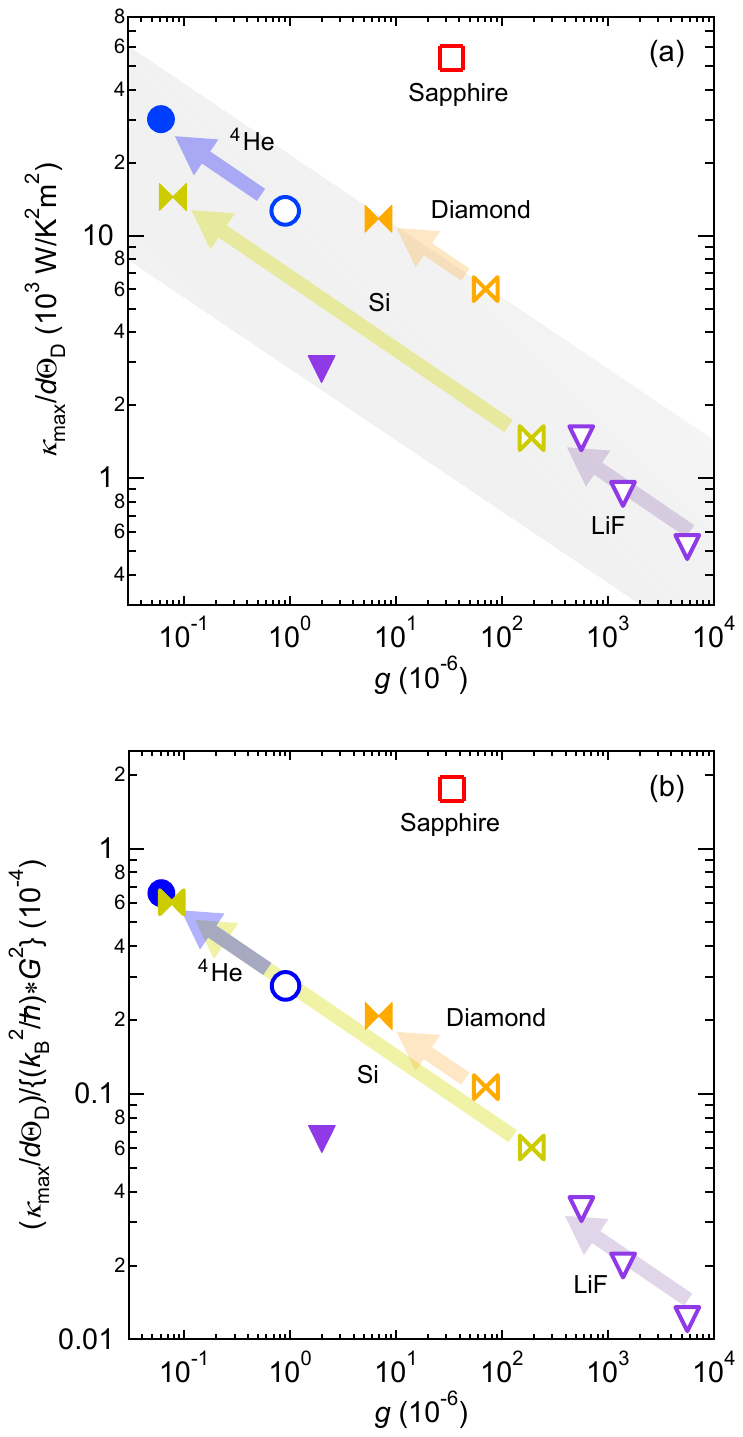}
\vspace*{-3.5cm} 
\caption{(a) A plot of the maximum thermal conductivity $\kappa_{\rm max}$ divided by the sample dimension $d$ and the Debye temperature $\Theta_{\rm D}$ against the isotope disorder parameter $g$. In each system, $\kappa_{\rm max}/d\Theta_{\rm D}$ for the isotopically pure and the natural abundant samples are presented by solid and open symbols, respectively.
For LiF, the data of several variations of isotope compositions are shown by open triangles. 
(b) A plot of the dimensionless thermal conductivity $(\kappa_{\rm max}/d\Theta_{\rm D})/\{(k_{\rm B}^2/\hbar)G^2\}$ vs. $g$.}
\label{isotope}
\end{center}
\end{figure}

The peculiar features of $\kappa(T)$ can be seen by an examination of the phonon mean-free-path $l_{\rm ph}$ extracted from the thermal conductivity, the specific heat $C$ and the phonon velocity $v_{\rm s}$ ($l_{\rm ph}=\frac{3\kappa}{v_{\rm s}C}$). Its evolution with temperature is shown in Fig.~\ref{kappa}(g). An exponential increase of $l_{\rm ph}$ ends up attaining a magnitude close to the sample dimensions (the horizontal dashed line in the inset of Fig.~\ref{kappa}(g)). At low temperature,  $l_{\rm ph}$ displays a local peak (around 30 K) and a local minimum (around 10 K) as commonly observed in systems where hydrodynamic phonon transport becomes dominant~\cite{Kopylov,Martelli,MachidaBP,MachidaC,Jaoui2022}.
In the miliKelvin temperature, $l_{\rm ph}$ saturates to a length comparable to sample dimensions as, expected in the ballistic regime (see the inset of Fig.~\ref{kappa}(g)). 

Let us put under scrutiny the exponential behavior $\kappa\propto\exp(A/T)$ in the Ziman regime. The Arrhenius plot shown in Fig.~\ref{kappa}(f) yields  $A$ = 337 K. The Debye temperature is $\Theta_{\rm D}\sim$ 1047 K, and therefore : $A\sim\Theta_{\rm D}/3$, as predicted by Peierls~\cite{Peierls}. One can see from Fig.~\ref{kappa}(e) that $A$ and the crossover temperature from the kinetic to Ziman regime are comparable to each other. 

Hydrodynamic regimes in sapphire are destroyed in presence of a significant population of point-like defects. This is shown by  comparison with a ruby sample purchased from Orbray corporation, in which about 3~\% of Al atoms are substituted by Cr atoms. As seen from Figs.~\ref{kappa}(e-g), the exponential increase, the faster than $T^3$ decrease, and the non-monotonic mean free path  are all absent in the ruby sample. As a result, the maximum $\kappa$ is damped by a factor of 50. The phonon mean-free-path $l_{\rm ph}$ smoothly increases all the way down to 3 K, approaching a value close to the sample dimension (represented by the solid horizontal line in the inset of Figs.~\ref{kappa}(g)) without showing any trace of the phonon hydrodynamics. In other words, a defect concentration of 3~\%  does not affect the kinetic and the ballistic regimes, but wipes out the Ziman and the Poiseuille regimes. 

Let us compare sapphire with other high-purity insulators. Although, this has remained unnoticed, isotopically enriched diamond~\cite{Wei}, Si~\cite{Inyushkin}, and LiF~\cite{Thacher} display a similar  exponential behavior, asymptotically evolving to a $1/T$ dependence at a temperature close to $A$. This is shown in Figs.~\ref{kappaval}(a)-(c)). This is also true of NaF~\cite{Slacknaf,Jackson} and CsI~\cite{Cahillcsi}, as seen in Fig. ~\ref{kappaval}(d) and (e). In all these cases, $\kappa$  deviates from the $1/T$ dependence (dashed line) around $T\sim A$, extracted from an Arrhenius fit, and asymptotically follows the exponential temperature dependence. Finally, as seen in Fig.~\ref{kappaval}(f), this is also the case of the two helium isotopes. $^4$He~\cite{Lawson} and $^3$He~\cite{Thomlinson1972}, both display a similar behavior. However, in both cases, the kinetic regime has not been attained.

Fig.~\ref{A} reveals universal feature common to high-purity insulators. The amplitude of $A$ as a function of $\Theta_{\rm D}$ in different systems is shown in Fig.~\ref{A}(a). One can see that in all of them, it lies between $\Theta_{\rm D}/5$ and $\Theta_{\rm D}/2$. 

The upper bound to $A$ can be justified considering Umklapp phonon-phonon collisions. They are only possible if the wave-vector of each colliding phonon, $k$, exceeds $G/4$, where $G=2\pi/a$ ($a$: the lattice constant) is a reciprocal lattice vector. Since $k=k_{\rm B}T/\hbar v_{\rm s}$, the energy of such a phonon is given by $k_{\rm B}T_{\rm U}=\pi\hbar v_{\rm s}/2a$. Assuming linear dispersion ($\omega_{\rm D}=k_{\rm B}\Theta_{\rm D}\sim v_{\rm s}\hbar(G/2)$), one finds $T_{\rm U}\sim\Theta_{\rm D}/2$. The average number of such a phonon decreases exponentially with temperature, $\langle N\rangle\sim\exp(-T_{\rm U}/T)\sim\exp(-\Theta_{\rm D}/2T)$, so that the thermal conductivity is $\kappa\propto\exp(\Theta_{\rm D}/2T)$. This leads to $A=\Theta_{\rm D}/2$ in good agreement with what is seen in Fig.~\ref{A}(a). 

It is instructive to consider exceptions to this correlation.  For example, in SrTiO$_3$, $A=$ 20~K ($\sim\Theta_{\rm D}/25$) ~\cite{Martelli}. This unusually small value points to a role played by low-energy soft phonons of this quantum paraelectric and their coupling with acoustic modes \cite{fauque2022}. The case of bismuth, where $A= 14.3~K$ ($\sim\Theta_{\rm D}/8$)~\cite{Gourgout}, while less striking, points towards a similar direction.

A more striking similarity between ultra-pure insulators is shown in Fig.~\ref{A}(b). It displays  $\kappa$, normalized by its value at $T=A$ as a function of $T/A$. All data fall into a universal curve, which is exponential for $T/A\leq 1$ and evolves towards $1/T$, when $T/A\geq 1$. This figure indicates that in a perfect insulator with no disorder (and no soft phonons) knowing $A$ (which is a fraction of $\Theta_{\rm D}$) is sufficient to know thermal conductivity at any temperature above its peak value. 

As seen in Figs.~\ref{kappaval}(a)-(d), in all these high-purity insulators, the random mixture of isotopes or impurities pulls down $\kappa$. Not only the exponential variation but also the prominent peak in $\kappa$ are considerably suppressed in naturally abundant diamond~\cite{BermanC}, Si~\cite{BermanC}, and LiF~\cite{Thacher} as well as in impure NaF~\cite{Jackson}.

The case of sapphire is outstanding, because it exhibits an exponential behavior in absence of isotopic purity. As shown in Table~\ref{table}, the dimensionless parameter quantifying isotopic disorder, $g$, is exceptionally high for sapphire, compared to other insulators displaying an exponential rise.  Moreover, the specific heat of our samples display a Schottky anomaly indicating the presence  of magnetic impurities  (Fig.~\ref{specific}).  Despite this level of isotopic and chemical impurities in the range of 30 ppm (see Table~\ref{table}), our sapphire samples behave like ultra-pure insulators.

\begin{table*}[bt]
\caption{Comparison of crystal structure, number of atoms in the unit cell $N$, isotope composition, dimensionless parameter of isotopic disorder $g$, Debye temperature $\Theta_{\rm D}$, and $A$ for materials exhibiting the exponential rise in thermal conductivity. Isotope composition of isotopically purified diamond, Si, LiF, and $^3$He is from Refs.~\cite{Wei,Inyushkin,Thacher,Thomlinson1972}.}
\newlength{\myheight}
\setlength{\myheight}{0.5cm}
\begin{tabular}{lccc|cc|cc|cc}
\parbox[c][\myheight][c]{0cm}{}
&&&&\multicolumn{2}{c|}{Natural abundant sample}&\multicolumn{2}{c|}{Isotopically pure sample}&&\\
\midrule[0.1pt]
\parbox[c][\myheight][c]{0cm}{}
&Crystal structure&$N$&Isotope&Composition (\%)&$g$ ($\times$10$^{-6}$)&Composition (\%)&$g$ ($\times$10$^{-6}$)&$\Theta_{\rm D}$ (K)&$A$ (K)\\
\midrule[0.1pt]
\parbox[c][\myheight][c]{0cm}{}
\multirow{4}{*}{Sapphire}&\multirow{4}{*}{Trigonal}&\multirow{4}{*}{30}&$^{27}$Al&100&\multirow{4}{*}{34}&-&\multirow{4}{*}{-}&\multirow{4}{*}{1047}&\multirow{4}{*}{337}\\
&&&$^{16}$O&99.76&&-&&&\\
&&&$^{17}$O&0.04&&-&&&\\
&&&$^{18}$O&0.21&&-&&&\\
\midrule[0.1pt]
\parbox[c][\myheight][c]{1cm}{}
\multirow{2}{*}{Diamond}&\multirow{2}{*}{Cubic (fcc)}&\multirow{2}{*}{8}&$^{12}$C&98.9&\multirow{2}{*}{74}&99.9&\multirow{2}{*}{7}&\multirow{2}{*}{2220}&\multirow{2}{*}{425}\\
&&&$^{13}$C&1.1&&0.1&&&\\
\midrule[0.1pt]
\parbox[c][\myheight][c]{0cm}{}
\multirow{3}{*}{Si}&\multirow{3}{*}{Cubic (fcc)}&\multirow{3}{*}{8}&$^{28}$Si&92.2&\multirow{3}{*}{200}&99.995&\multirow{3}{*}{0.08}&\multirow{3}{*}{645}&\multirow{3}{*}{331}\\
&&&$^{29}$Si&4.7&&0.0046&&&\\
&&&$^{30}$Si&3.1&&0.0004&&&\\
\midrule[0.1pt]
\parbox[c][\myheight][c]{0cm}{}
\multirow{3}{*}{LiF}&\multirow{3}{*}{Cubic (Rock salt)}&\multirow{3}{*}{8}&$^{6}$Li&7.6&\multirow{3}{*}{1460}&0.01&\multirow{3}{*}{2}&\multirow{3}{*}{735}&\multirow{3}{*}{184}\\
&&&$^{7}$Li&92.4&&99.99&&&\\
&&&$^{19}$F&100&&100&&&\\
\midrule[0.1pt]
\parbox[c][\myheight][c]{0cm}{}
\multirow{2}{*}{NaF}&\multirow{2}{*}{Cubic (Rock salt)}&\multirow{2}{*}{8}&$^{23}$Na&100&\multirow{2}{*}{0}&-&\multirow{2}{*}{-}&\multirow{2}{*}{488}&\multirow{2}{*}{128}\\
&&&$^{19}$F&100&&-&&&\\
\midrule[0.1pt]
\parbox[c][\myheight][c]{0cm}{}
\multirow{2}{*}{CsI}&\multirow{2}{*}{Cubic (bcc)}&\multirow{2}{*}{2}&$^{133}$Cs&100&\multirow{2}{*}{0}&-&\multirow{2}{*}{-}&\multirow{2}{*}{128}&\multirow{2}{*}{34}\\
&&&$^{127}$I&100&&-&&&\\
\midrule[0.1pt]
\parbox[c][\myheight][c]{0cm}{}
\multirow{2}{*}{$^4$He}&\multirow{2}{*}{Hexagonal}&\multirow{2}{*}{2}&$^{3}$He&0.0001&\multirow{2}{*}{0.06}&-&\multirow{2}{*}{-}&\multirow{2}{*}{36.7}&\multirow{2}{*}{7.3}\\
&&&$^{4}$He&99.9999&&-&&&\\
\midrule[0.1pt]
\parbox[c][\myheight][c]{0cm}{}
\multirow{2}{*}{$^3$He}&\multirow{2}{*}{Cubic (bcc)}&\multirow{2}{*}{2}&$^{3}$He&-&\multirow{2}{*}{-}&99.9998&\multirow{2}{*}{0.2}&\multirow{2}{*}{23}&\multirow{2}{*}{4.5}\\
&&&$^{4}$He&-&&0.0002&&&\\
\bottomrule[0.1pt]
\label{table}
\end{tabular}
\end{table*}

The exceptional case of sapphire is emphasized by considering the amplitude of the peak thermal conductivity $\kappa_{\rm max}$ as a function of $g$. The magnitude of $\kappa_{\rm max}$ depends on the sample dimension, $d$, and the Debye temperature, $\Theta_{\rm D}$. Increasing them would enhance $\kappa_{\rm max}$, by allowing the exponential increase to persist down to lower temperatures. Fig.~\ref{isotope}(a) is a plot of $\kappa_{\rm max}$ normalized by a product of $d \times \Theta_{\rm D}$  as a function of $g$. 
The isotopic purification pushes up $\kappa_{\rm max}/d\Theta_{\rm D}$ towards higher values (from the open to closed symbol in each material) following $\kappa_{\rm max}/d\Theta_{\rm D}\propto g^{-0.3}$ as guided by the solid arrows. 
As seen in  the figure, $\kappa_{\rm max}/d\Theta_{\rm D}$ in sapphire stands out. Despite a $g$ 500 times larger, sapphire surpasses its close rival, $^4$He.

Let us now consider a dimensionless thermal conductivity, the ratio of $\kappa_{\rm max}/d\Theta_{\rm D}$ to $(k_{\rm B}^2/\hbar)G^2$, where $G$ is the width of Brillouin zone. As seen in Fig.~\ref{isotope}(b), the available data collapse onto a single line, implying that the effect of isotope scattering on the phonon heat transport across a variety of insulating solids is properly accounted by $g$. One can see again that sapphire does not participate in this scaling, indicating a puzzling robustness of its high thermal conductivity against the isotopes. 

Compared to other insulators discussed here, sapphire has many more atoms, $N$, in its primitive cell. As listed in Table~1, $N=30$ for sapphire, $N=8$ for diamond, Si, and LiF with fcc or rock salt structure, and $N=2$ for hexagonal solid $^4$H.  A large $N$ attenuates mass perturbation  by isotopic impurities. Moreover, it leads to an increase in the number of optical phonon branches. As shown in Fig.~\ref{dispersion}(a), in sapphire the gap between the optical and the acoustic branches is small. This enhances the phase space for scattering between the acoustic and optical phonons and reduces a chance for the acoustic phonons to be scattered by the isotopes. Notably, the fact that the energy range ($\sim$10~THz) where the high frequency acoustic phonons can meet the optical phonons is close to the upper bound of Ziman regime, that is $A$ ($\sim$7~THz), suggests that both phonon modes indeed participate the Normal processes. Because of the proximity of optical and acoustic phonon energies, the thermal conductivity of sapphire, in contrast to other `simple' insulators, is not set by $g$.   

We note that this scenario is compatible with theoretical calculations suggesting that isotopic purity does not necessarily lead to  higher thermal conductivity~\cite{Lindsay}. It has been demonstrated that isotopic purification of natural LiH by replacing $^7$Li with $^6$Li leads to a higher frequency of acoustic modes and a smaller phonon gap. This allows for increased coupling of the heat-carrying acoustic phonons with the optical phonons and reduction of their lifetimes. As a result, thermal conductivity of the isotopically pure $^6$LiH is lower than that of the natural LiH. The possible correlation between isotope mass variations and Normal collision rate emerges as a potential theme for future theoretical studies.

In this context, let us also recall the case of another highly conducting oxide, In$_2$O$_3$, which has forty atoms per primitive cell and numerous optical phonons and its thermal conductivity can surpass that of naturally abundant silicon \cite{Xu2021}. 

To sum up, naturally occurring sapphire displays the two phonon hydrodynamic regimes and a thermal conductivity much higher than what is expected, given its isotopic purity. We identify the large number of atoms per the primitive cell as a possible source of frequent momentum exchange among phonons including optical modes and attenuation of phonon-isotope scattering.

We acknowledge Kentaro Kitagawa, Laurent Jalabert, and Masahiro Nomura for technical supports. This work was supported by Grants-in-Aid for Scientific Research
(KAKENHI Grants No. JP19H01840 and No. JP23H00092).


\setcounter{figure}{0}
\renewcommand{\thefigure}{S\arabic{figure}}
\section{APPENDIX}
\subsection{Thermal conductivity measurements}
Thermal conductivity data below 300 K was obtained by employing the steady-state method.
Since the radiation of heat from the sample surface becomes a serious problem at high temperatures, we employed the 3$\omega$ method~\cite{Cahill} above 80~K.
In the 3$\omega$ experiment, the sample is heated locally so that compared to steady-state experiments, errors due to radiation are reduced to a negligibly low level.
The single crystals with a size of 0.5 $\times$ 0.6 $\times$ 10 mm$^3$ and 5 $\times$ 5 $\times$ 3 mm$^3$ were used for the steady-state and 3$\omega$ method, respectively.

\subsubsection{Steady-state method}
The steady-state experiments by the standard one-heater-two-thermometers method were performed in the commercial measurement system (Quantum Design PPMS) between 2 and 300~K and in the home-made dilution refrigerator between 0.1 and 2.5~K.
The measurements were carried out in a high vacuum level better than 10$^{-4}$~Pa.
The thermal gradient in the sample was produced through a chip resistor alimented by a current source (Keithley 6220). 
The DC voltage on the heater was measured by the digital multimeter (Keithley 2000).
The thermometers, heater, and cold finger were connected to samples directly or by gold wires of 25~$\mu$m diameter. 
All contacts on the sample were made using silver paste Dupont 4922N. 
The heat current was injected along the largest length of the sample (10~mm) which is perpendicular to the $c$-axis.
The temperature difference generated in the sample was determined by measuring the local temperature with two thermometers: E-type thermocouple (5~K-300~K) and Cernox resistors (2~K-15~K) in the PPMS, and RuO$_2$ resistors (0.1~K-2.5~K) in the dilution refrigerator.
The DC voltage on the thermocouple was measured by the DC-nanovoltmeter (Keithley 2182A).
The resistance of Cernox and RuO$_2$ thermometers were measured by the resistance bridges (Cryocon 24C) and the MMR3 resistance bridge (Cryoconcept), respectively.

The heat loss along manganin wires connected to the thermometers and heater is many orders of magnitude lower than the thermal current along the sample. 
Because of the large thermal conductivity $\kappa\sim$~35,000~W/Km around the peak and the tiny temperature difference in the sample, which is as small as a few millikelvin, the scattering of the data is large. Thus, the measurements were repeated several times and the data is averaged.
Error bars in the main figures represent one standard deviation.

\subsubsection{3$\omega$ method}
\begin{figure}[tb]
\begin{center}
\centering
\includegraphics[width=11cm]{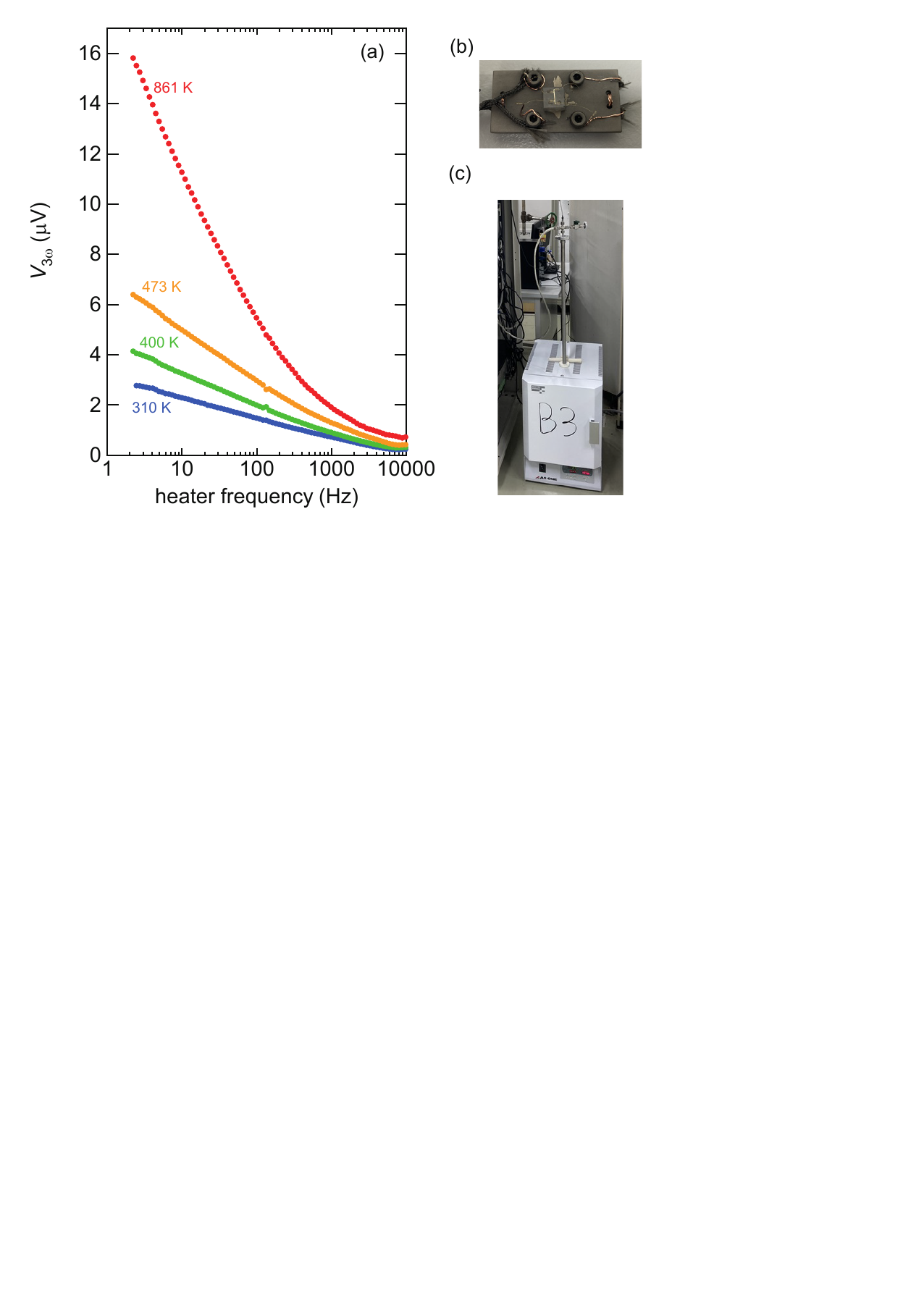}
\vspace*{-9.5cm} 
\caption{(a) Heating frequency dependence of the in-phase 3$\omega$ voltage oscillation $V_{\rm 3\omega}$ collected at different temperatures. Photos of the setup for the 3$\omega$ measurements at high temperature (b) and the custom-made probe inserted in the electric furnace (c).}
\label{3omega}
\end{center}
\end{figure}

The 3$\omega$ experiments were performed in the $N_2$ cryostat between 80 and 350~K and in the electric furnace between 300 and 900~K.
The narrow gold line serving as both the heater and the thermometer had a width of 100~$\mu$m, a thickness of 70~nm, and a length of 3~mm was directly 
sputtered on the surface of sapphire sample.
The resistance of this line is $\sim$ 5~$\Omega$ at 300 K.
The line structure shown in Fig.~\ref{kappa}(d) and Fig.~\ref{3omega} (b) were made by using a thin stainless mask.
A sinusoidal current with a typical amplitude of 20~mA generated by a current source (Keithley 6221) is passed through the line with angular frequency $\omega$, which generates power varying as 2$\omega$.
This power creates thermal oscillations varying as 2$\omega$.
Due to the linear temperature dependence of the metallic heater, the 2$\omega$ thermal oscillation causes 
the resistance of the line to vary as 2$\omega$. 
This resistance oscillation mixes with the input current and generates a component
of the measured voltage which varies as 3$\omega$.

Applying the 3$\omega$ method to bulk geometry, the in-phase 3$\omega$ voltage oscillation $V_{\rm 3\omega}$ of the line is expected to be linear in logarithmic heating frequency 2$f$ ($f=\omega/2\pi$) as long as the thermal penetration depth is large compared to the heater half width $b\sim$ 50~$\mu$m and at least five times smaller than the sample thickness $t$ = 3 mm.
The gold line was connected in series with a programmable resistor (IET PRS-330), which is adjusted to have the same resistance with the line.
Home-made unity gain differential amplifiers were employed to acquire the voltage signal across the programmable resistor and the sample circuit.
$V_{\rm 3\omega}$ was obtained by subtracting the first harmonic voltage across the programmable resistor ($V_{\rm 1\omega}$) from that across the line ($V_{\rm 1\omega+\rm 3\omega}$).
By measuring $V_{\rm 3\omega}$ as a function of frequency with a lock-in amplifier (Signal Recovery 7265), the
thermal conductivity $\kappa$ of the material can be extracted from the slope of $V_{\rm 3\omega}$ vs. log2$f$.
A typical frequency scan at different temperatures is shown in Fig.~\ref{3omega}(a).

For the 3$\omega$ measurements above 300~K, the sample was mounted on the alumina plate (Fig.~\ref{3omega}(b)) by silver paste and set in a custom-made probe with the K-type thermocouple for the measurement of sample temperature. Inside of the probe was kept under a vacuum of 10$^{-4}$~Pa.
The probe was inserted in the electrical furnace which can control temperature in a range between 100 and 1280 degrees Celsius (Fig.~\ref{3omega}(c)).
The measurements in the $N_2$ cryostat was carried out with a custom-made probe under a vacuum of 10$^{-4}$~Pa. 
The sample was mounted on the copper block by GE varnish and set in the probe.
The probe was dipped in liquid nitrogen and the sample temperature was regulated by the temperature controller (Cryocon 24C) by monitoring the Pt thermometer attached close to the sample.

\subsection{Reproducibility}
\begin{figure}[tb]
\begin{center}
\centering
\includegraphics[width=8cm]{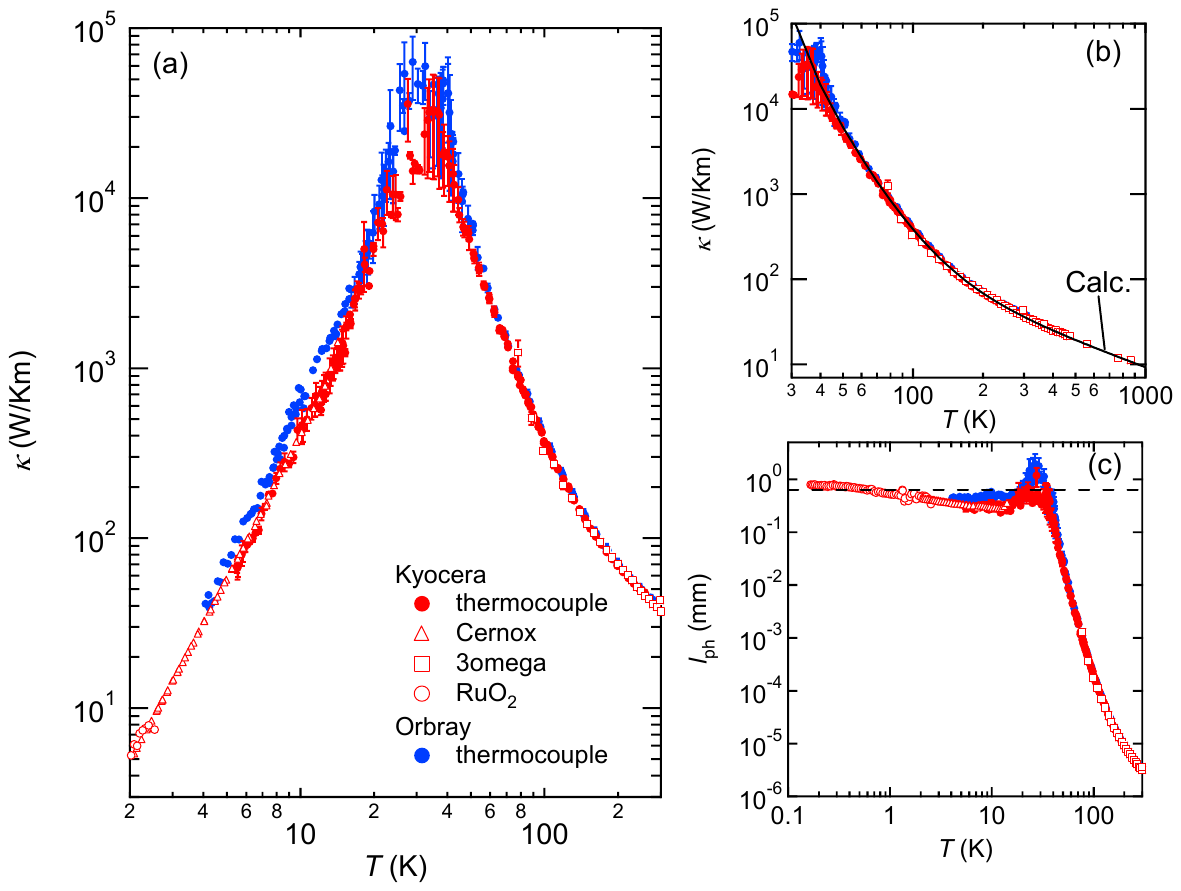}
\vspace*{-5.5cm} 
\caption{(a) A comparison of thermal conductivity of sapphire samples from two different sources. One purchased from Kyocera corporation and the other from Orbray corporation. (b) Our results well reproduce the first principle calculation~\cite{subedi21} in the Ziman and kinetic regimes. (c) Temperature variations of phonon mean-free-path of the two samples.}
\label{kyoob}
\end{center}
\end{figure}

We have measured thermal conductivity of the different sapphire sample purchased from the different company (Orbray corporation) to check the reproducibility.
As can be seen from Figs.~\ref{kyoob}(a) and (b), $\kappa$ from the two different sources well coincides each other. Our results also reproduce the first-principle calculations~\cite{subedi21} in the Ziman and kinetic regimes as shown in Fig.~\ref{kyoob}(b). Since the two samples have the comparable sample dimension ($d\sim$ 0.62 mm and 0.67 mm for the Kyocera and Orbray samples, respectively), the phonon mean-fee-path $l_{\rm ph}$ saturates to almost the same value at low temperature (Fig.~\ref{kyoob}(c)). On warming, $l_{\rm ph}$ for the both samples causes the minimum around 10 K, then peaks around 30 K. At the maximum, $l_{\rm ph}$ exceeds $d$ as expected for the Poiseuille flow of phonons.

\subsection{Comparison to the literature data}
\begin{figure}[tb]
\begin{center}
\centering
\includegraphics[width=8cm]{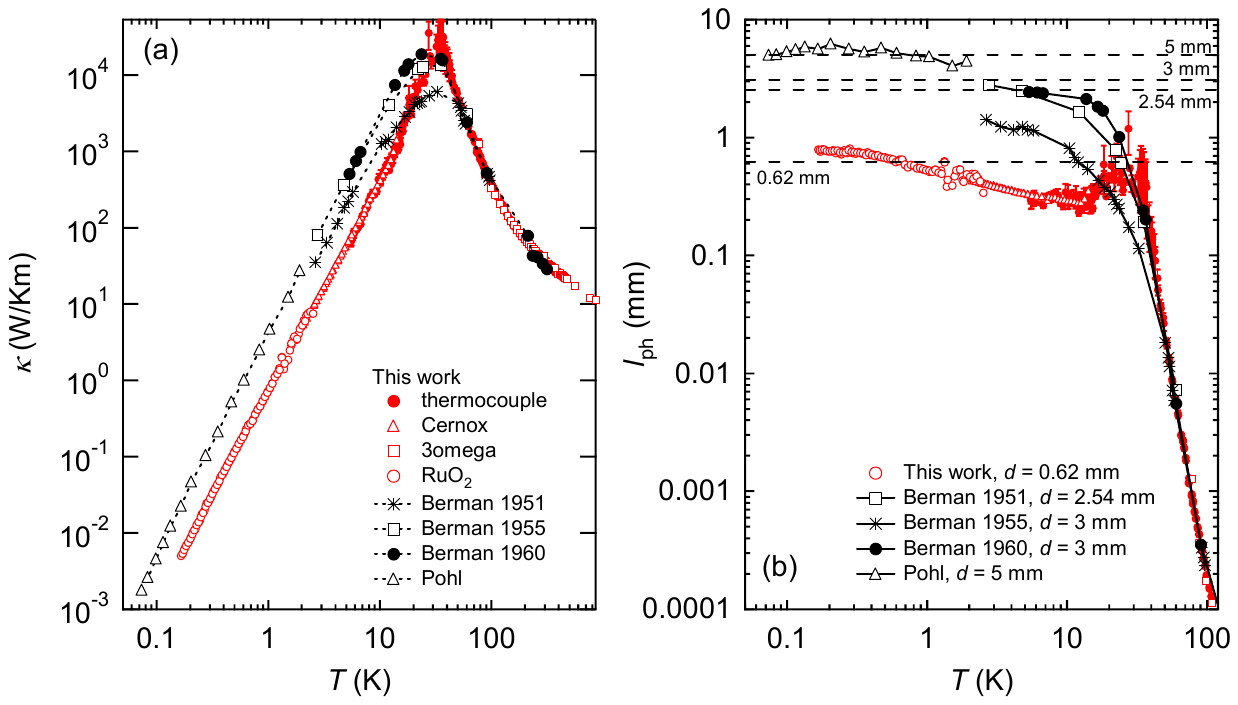}
\vspace*{-7cm} 
\caption{A comparison of (a) thermal conductivity and (b) phonon mean-fee-path data with the previous reports by Berman~\cite{Berman1951,Berman1955,Berman1960} and Pohl~\cite{Pohl}.}
\label{comp}
\end{center}
\end{figure}

In Figure~\ref{comp}(a), thermal conductivity of the Kyocera sample is shown together
with the previously reported data by Berman~\cite{Berman1951,Berman1955,Berman1960} and Pohl~\cite{Pohl}. Our data reproduces the exponential behavior of Berman's data in the Ziman regime and the $T^3$ dependence of Pohl's data in the ballistic regime ($T<$ 1 K).
The larger magnitude of Pohl's data is due to the larger sample dimension $d=$ 5 mm.
Figure~\ref{comp}(b) depicts temperature variation of $l_{\rm ph}$ extracted from $\kappa$ of the previous reports. One can see that $l_{\rm ph}$ shows a tendency to saturate to each $d$.

In the literature data, the Poiseuille regime is not detectable probably due to the low quality of samples fabricated in many decades ago, implying that the Poiseuille flow is more fragile against impurities and defects compared to the Ziman hydrodynamics.

\begin{figure}[tb]
\begin{center}
\centering
\includegraphics[width=9cm]{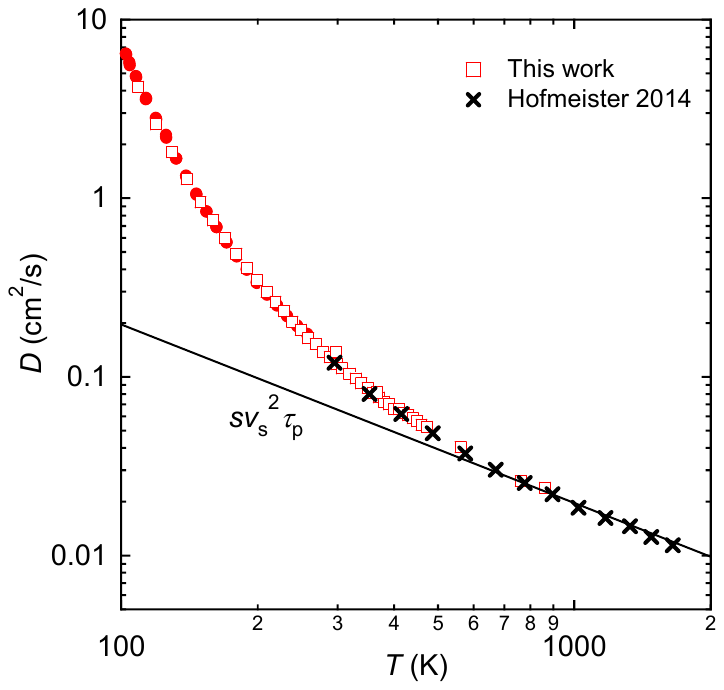}
\vspace*{-8cm} 
\caption{A comparison of temperature dependence of thermal diffusivity $D$ with the literature data~\cite{Hofmeister}. The solid line represents $sv_{\rm s}^2\tau$ with $s$ = 2.04 ~\cite{Mousatov} and $v_{\rm s}$ = 11.23 km.}
\label{diffusivity}
\end{center}
\end{figure}

Figure~\ref{diffusivity} shows temperature dependence of thermal diffusivity $D$, which is the
ratio of thermal conductivity to specific heat $C$, $D=\kappa/C$.
Our results agree with the previous report~\cite{Hofmeister}. Above 600 K, $D$ follows $1/T$ dependence and respect the Plankian bound $D\geq sv_{\rm s}^2\tau_{\rm p}$ ($\tau_{\rm p}=\hbar/k_{\rm B}T$)~\cite{Behniapl,Mousatov} with $v_{\rm s}=11.23$ km/s.
Here, $s>$ 1 is a constant specific for different families of materials.
We employed $s=2.04$ as suggested by Ref.~\cite{Mousatov}.

\subsection{Specific heat}
\begin{figure}[tb]
\begin{center}
\centering
\includegraphics[width=9cm]{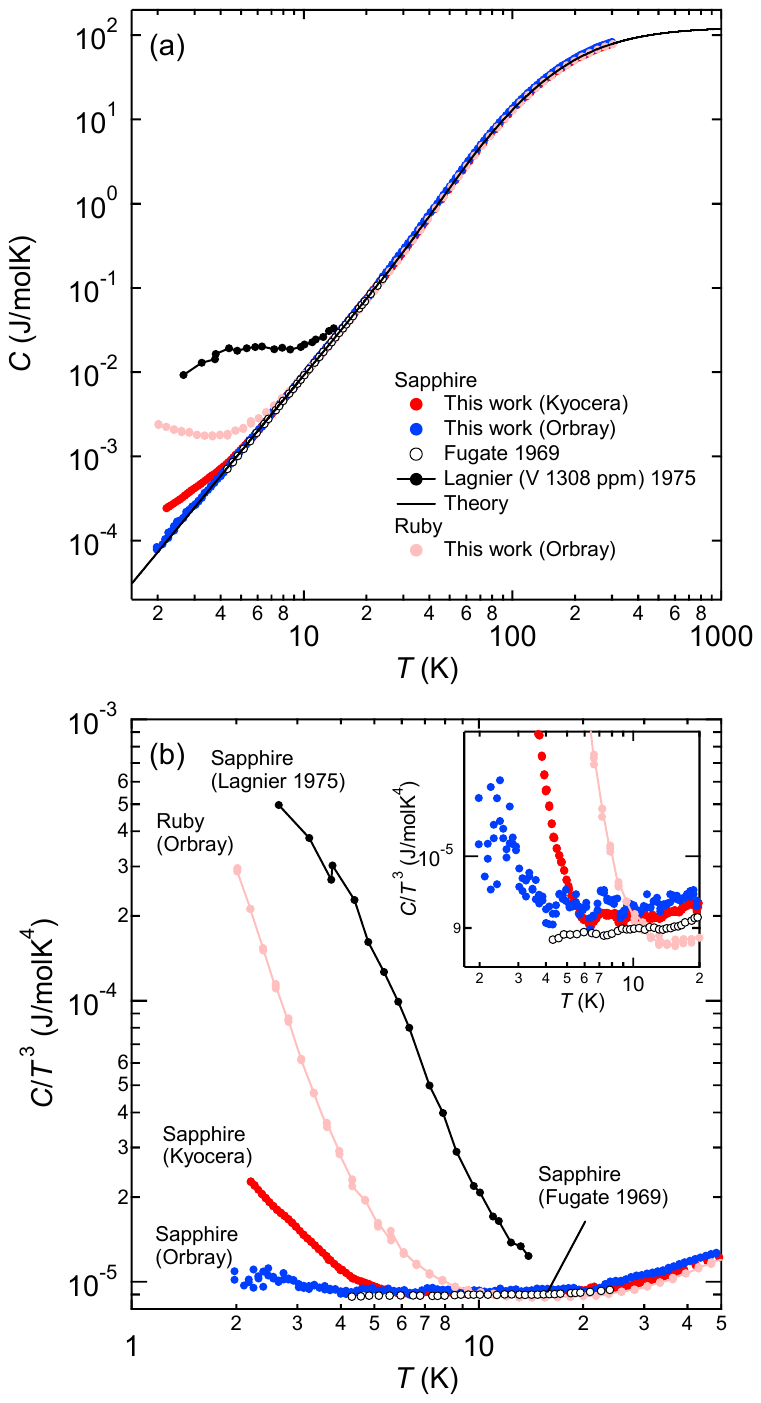}
\vspace*{-2.5cm} 
\caption{A comparison of temperature dependence of specific heat $C$ (a) and $C/T^3$ (b) of the sapphire and ruby samples together with the literature data~\cite{Fugate,Lagnier}. The theoretical calculation is also shown in panel (a).
As shown in the panel (b) and its inset, the excess specific heat as indicated by an upward deviation from the $T^3$ dependence is present in all our samples.}
\label{specific}
\end{center}
\end{figure}

Specific heat $C$ of single-crystalline sapphire and ruby was measured between 2 K and 300 K using a relaxation method of the heat capacity option in a Quantum Design, PPMS instrument. The sapphire samples purchased from Kyocera and Orbray corporation and the ruby sample from Orbray corporation were used. 
As shown in Fig.~\ref{specific}(a), our results of sapphire agree well with the previous report~\cite{Fugate} and the first-principle calculation except for upturns at low temperature. 
The upturn causes a deviation from the $T^3$ Debye specific heat as shown in Fig.~\ref{specific}(b) where $C$ divided by $T^3$ is plotted as a function of temperature.
The similar upturn was reported in Ref.~\cite{Lagnier} where the excess component is attributed to the Schottky specific heat  due to 1308 ppm vanadium contamination.
We infer that our specific heat data is also affected by the magnetic impurities which are chromium for the ruby sample, but unknown for the sapphire samples.

For the estimation of phonon mean-free-path, we neglect the upturn and assume a constant $C/T^3$ down to low temperature, which is $9.15\times 10^{-6}$~J/molK$^4$, $9.33\times 10^{-6}$~J/molK$^4$, and $8.86\times 10^{-6}$~J/molK$^4$ for the sapphire samples from Kyocera and Orbray, and the ruby sample, respectively. Above 300~K where the experimental data of specific heat is not available, we use the calculated results (solid line in Fig.~\ref{specific}(a)) to evaluate phonon mean-free-path.

\subsection{Computational Method}
The phonon contribution to the specific heat was calculated within the quasiharmonic
approximation using the {\sc shengbte} code \cite{shengbte}.  This requires the calculation of 
second-order interatomic force constants (IFC), which was obtained using the supercell-based 
frozen phonon approach as implemented in the {\sc phonopy} pacakge \cite{phonopy}.  We used $4 \times
4 \times 4$ supercells to generate the necessary displaced structures, and the atomic forces in 
these structures were obtained from first principles using the {\sc vasp} code within the 
local density approximation (LDA) \cite{vasp}.  These calculations were performed using a plane-wave 
cutoff of 520 eV and $k$-point mesh of $3 \times 3 \times 3$.  The IFCs were obtained using 
the fully-relaxed structure minimizing both the lattice stresses and atomic forces.  The phonon 
density of states used in the calculation of the specific heat was obtained by interpolating the 
IFCs on an $108 \times 108 \times 108$ $q$-point mesh. 

\subsection{Comparison to the other insulators}
\begin{figure*}[tb]
\begin{center}
\centering
\includegraphics[width=16cm]{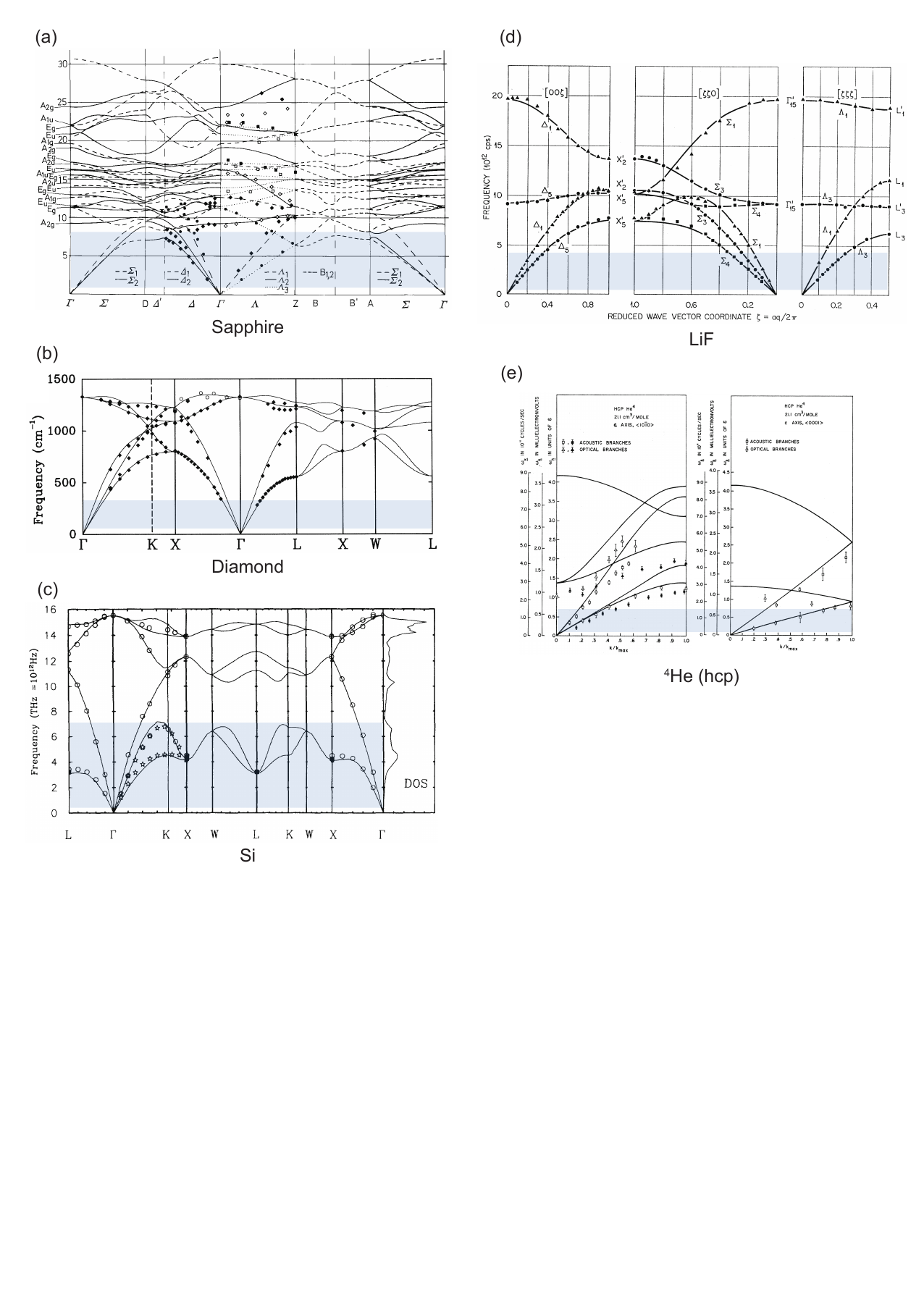}
\vspace*{-7cm} 
\caption{A comparison of phonon dispersion of (a) sapphire~\cite{KappusAl2O3}, (b) diamond~\cite{PavoneC}, (c) Si~\cite{WeiSi}, (d) LiF~\cite{DollingLiF}, and (e) hcp $^4$He~\cite{GillisHe}. Shaded region represents the energy-range corresponding to the Ziman regime which is between the conductivity maximum temperature and $A$.}
\label{dispersion}
\end{center}
\end{figure*}

In Table~1, we compare sapphire with the other insulators showing the Ziman heat transport.
There are two things to be emphasized.
(i) Dimensionless parameter of isotopic disorder $g$ for sapphire is exceptionally high.
(ii) Sapphire contains a large number of atoms in the unit cell.
Related to the latter feature, there are a lot of the optical phonon modes in sapphire as shown in Fig.~\ref{dispersion}(a).
This is contrasting with fewer optical modes in diamond, Si, LiF, and $^4$He (Figs.~\ref{dispersion}(b)-(e)) in which a number of atoms in the unit cell is much smaller (see Table~1).
The large number of optical modes combined with a small gap between the optical and acoustic branches allows coupling between of heat-carrying acoustic phonons with the optical phonons.
This weakens the phonon-isotope scattering, but facilitates the intrinsic phonon-phonon scattering.
The fact that the Ziman regime (shaded region in Figs.~\ref{dispersion}(a)-(e)) extends up to the energy range where the acoustic phonons can meet the optical phonons indicates that both phonon modes participate the Ziman transport in sapphire. In the other insulators, the only acoustic phonons are active in the corresponding regime.

\bibliography{ref}

\end{document}